\documentclass[prb, aps, reprint, twocolumn, showpacs, preprintnumbers, amsmath, amssymb]{revtex4-1}

\usepackage{graphicx}
\usepackage{dcolumn}
\usepackage{bm}
\usepackage{epstopdf}

\begin{document}
\preprint{APS/123-QED}

\title{Dissipation signatures of the normal and superfluid phases in torsion pendulum experiments with $^3$He in aerogel}
\author{N. Zhelev$^{a}$, R.G. Bennett$^{a}$, E.N. Smith$^{a}$, J. Pollanen$^{b}$,  W.P. Halperin $^{b}$, and J.M. Parpia$^{a}$}
\affiliation{$^{a}$Laboratory of Atomic \& Solid State Physics, Cornell University, Ithaca, NY, 14853 USA}
\affiliation{$^{b}$Department of Physics and Astronomy, Northwestern University, Evanston, IL 60208, USA}

\date{\today}
\begin{abstract}

We present data for energy dissipation factor ($Q^{-1}$) over a broad temperature range at various pressures of a torsion pendulum setup used to study $^3$He confined in a 98\% open silica aerogel.
Values for $Q^{-1}$ above $T_c$ are temperature independent and have weak pressure dependence.
Below $T_c$, a deliberate axial compression of the aerogel by 10\% widens the range of metastability for a superfluid Equal Spin Pairing (ESP) state; we observe this ESP phase on cooling and the B phase on warming over an extended temperature region.
While the dissipation for the B phase tends to zero as T $\rightarrow$ 0, $Q^{-1}$ exhibits a peak value greater than that at $T_c$ at intermediate temperatures.
Values for $Q^{-1}$ in the ESP phase are consistently higher than in the B phase and are proportional to $\rho_s/\rho$ until the ESP to B phase transition is attained.
We apply a viscoelastic collision-drag model, which couples the motion of the helium and the aerogel through a frictional relaxation time $\tau_f$. Our dissipation data is not sensitive to the damping due to the presumed small but non-zero value of $\tau_f$. The result is that an additional mechanism to dissipate energy not captured in the collision-drag model and related to the emergence of the superfluid order must exist.
The extra dissipation below $T_c$ is possibly associated with mutual friction between the superfluid phases and the clamped normal fluid.
The pressure dependence of the measured dissipation in both superfluid phases is likely related to the pressure dependence of the gap structure of the ``dirty" superfluid.
The large dissipation in the ESP state is consistent with the phase being the A or the Polar with the order parameter nodes oriented in the plane of the cell and perpendicular to the aerogel anisotropy axis.

\end{abstract}


\maketitle

\section{Introduction}

Unconventionally paired Fermi systems exhibit strong sensitivity in their transport to the presence of even a small degree of nonmagnetic impurities.\cite{ulm, basov, bernhard, fujita, graser, sanna}
For the otherwise pure superfluid $^3$He, an elastic scattering mechanism, in addition to inelastic two-particle scattering processes, is provided by porous silica aerogel ``impurities."\cite{thuneberg1, hanninen1, sauls, einzel}
Since the discovery of superfluidity of $^3$He in aerogel,\cite{porto, halperin} the analogy of this so-called ``dirty" Fermi superfluid with ``dirty" unconventional superconductors has been investigated in the literature.
Transport measurements in the normal Fermi liquid (spin,\cite{grenoble, eska} thermal conductivity,\cite{lancasternormal, saulslt2000} and viscosity\cite{einzel}) reveal a crossover from an intrinsic inelastic quasiparticle-quasiparticle (qp-qp) scattering rate at high temperatures, to a quasiparticle-impurity dominated relaxation mechanism at a lower temperature.

In the $^3$He-aerogel composite system, the $^3$He is always on the order of the zero-temperature coherence length away from the aerogel strands. The zero-temperature coherence length is defined to be $\xi_0=\hslash v_F/2\pi k_BT_c$.
It is expected that the superfluid order parameter is suppressed and surface bound states exist near macroscopic surfaces and domain walls.\cite{NagatoJLTP, VorontsovSauls, TsutsumiMachida} However, the aerogel strands do not act as conventional surfaces -- else superfluidity would be entirely suppressed in the $^3$He-aerogel system. Instead, scattering from the aerogel leads to a suppression of the superfluid gap. We expect a spectrum of low energy excitations inside the gap to appear, which could lead to a gapless superfluid state in which the density of states is finite around the entire Fermi surface. \cite{sauls}
Evidence for such states exists in thermal conductivity\cite{lancaster_gapless} and heat capacity\cite{Choi_heatcap} measurements as $T\rightarrow0$, but the exact profile for the density of states of the $^3$He in aerogel system and its dependence on strong coupling effects is still not fully understood.

In order to probe the dynamics of the aerogel embedded fluid, we employ a torsion pendulum technique.
We track the frequency and quality factor ($Q$) of the pendulum with temperature.
Observing the frequency shift has proved instrumental in studying the effects of disorder at the onset of superfluid transition.\cite{porto, BennettZhelev}
However, due to the close spacing between the aerogel strands (of the order of $~50  $ nm), even the small impurity limited viscosity of the normal state $^3$He would be sufficient to clamp the fluid at the audio frequencies (2.1 kHz) corresponding to the driven antisymmetric torsional mode we employ.
In order to probe the transport properties, we cannot rely only on the frequency shift data.
Instead, in this article we focus on the energy dissipation factor ($Q^{-1}$) of the pendulum, which should also be sensitive to the Fermi surface excitations discussed in the previous paragraph.

The aerogel sample is deliberately compressed along the pendulum axis by 10\%.
It has been generally accepted that the aerogel anisotropy due to the axial compression should favor the anisotropic, equal spin pairing (ESP) superfluid $^3$He-A phase.\cite{Aoyama, Volovik} It had also been expected that the $\ell$ vector would tend to align along the axis of compression, however, recent pulsed NMR tip angle measurements on axially compressed aerogel at moderate magnetic fields (both along and perpendicular to the strain axis) show that the $\ell$ vector tends to be oriented in the plane of the cell and perpendicular to the strain axis regardless of the direction of the magnetic field.\cite{JiaLi}
Recent theoretical results\cite{Sauls2013} also point to the possibility of a Polar phase (also an ESP phase) with a line of nodes away from the strain axis.
In an earlier work we observed that the superfluid fraction in the ESP phase is less than that in the B phase.\cite{BennettZhelev}
If $\ell$ in the A phase (nodal direction in the Polar phase) was aligned perpendicular to the flow, we would instead observe the superfluid fraction in the A phase to exceed that in the B phase.\cite{Combescot,Berthold}
Thus either an A phase with $\ell$ randomly oriented along the plane of the cell or a Polar phase is consistent with the equal spin pairing state realized in this experiment. Lacking NMR data to identify the phase at zero magnetic field, we refer to the intervening phase as ESP rather than the A/Polar phase.

The metastable ESP phase we observe is supercooled to temperatures well below the equilibrium ESP to B phase boundary. On the other hand, after completion of the ESP to B transition by further cooling the cell, the superfluid B phase persists on warming and the ESP phase only reappears in a region of small temperature width very close to $T_c$. This results in a significant range of temperatures over which we have ESP phase on cooling and B phase on warming, and allows us to make a direct comparison of the properties ($\rho_s, Q^{-1}$) of the two superfluid phases.

In the following sections, we briefly outline experimental details, and present the experimental data.
Then we discuss a model for the energy dissipation factor of the torsion pendulum arising from the normal state fluid. Finally, we discuss the data below $T_c$, where we observe additional dissipation intrinsic to the superfluid. We relate our data to the presented theoretical model and propose other possible mechanism that could account for the observed behavior.

\section{Experimental Setup}

The aerogel was grown directly into a pillbox shaped stainless steel cavity consisting of a tightly fitted lid, a base and a shim inserted between them.
More about the physical properties and method of growth of aerogel can be found in Ref. \onlinecite{pollanen1}.

The aerogel was compressed by 10\% along its main axis by removing the shim and pressing the lid onto the base, bringing the height of the cell to 400  $\mu$m.
The height was chosen to be small enough to couple the aerogel well to the walls (though aerogel displacement relative to the cell's walls still needs to be considered), but large enough to ensure fine resolution in determining the fraction of superfluid.

The moment of inertia of the torsion head and aerogel filled cell was calculated to be 0.064 g-cm$^2$ and that of the helium at saturated vapor pressure -- 5.85$\times$10$^{-5}$ g-cm$^2$, or about 1 part in 10$^3$ of the inertia of the head. Our signal to noise ratio was $\approx$ 5$\times 10^{7}$, which provided an ample resolution in the data. Quality factor ($Q$) of the empty cell at low temperatures exceeded one million.

The steel cavity was dry fitted into an already hardened epoxy cast in order to reduce possible contamination of the aerogel by any epoxy penetrating through holes on the stainless pillbox.
Despite careful machining of the epoxy cast, there appeared to be empty regions around the periphery of the cell occupied by $^3$He not embedded in the aerogel (bulk fluid). In addition, the bulk fluid within the 1 mm diameter fill line needs to be considered. Appendix A describes how we modeled the contribution coming from these two regions. All the bulk fluid contribution has been subtracted from the data.

The pendulum was driven and detected capacitively.
All the measurements were performed while maintaining the pendulum at resonance using a phase-locked loop and adjusting the drive to keep a constant amplitude of motion.
The amplitude was chosen to be small enough so that any temperature dependent non-linear behavior could be avoided.
This meant that the antisymmetric mode of the torsion oscillator was operated at an amplitude of $\approx$ 0.1 nm, leading to a peak velocity of order 1 $\mu$m/sec. Dissipation ($Q^{-1}$) data was obtained from the ratio of the drive amplitude and the constant pendulum amplitude of motion.

More detailed discussion on the assembly of the cell along with a detailed plot for the empty cell background (period and dissipation ($Q^{-1}$)) can be found in Ref. \onlinecite{bennettjltp}.

\section{Data}

\begin{figure*}
\begin{center}
\includegraphics[%
  width=6.6in,
  height = 2.5in]{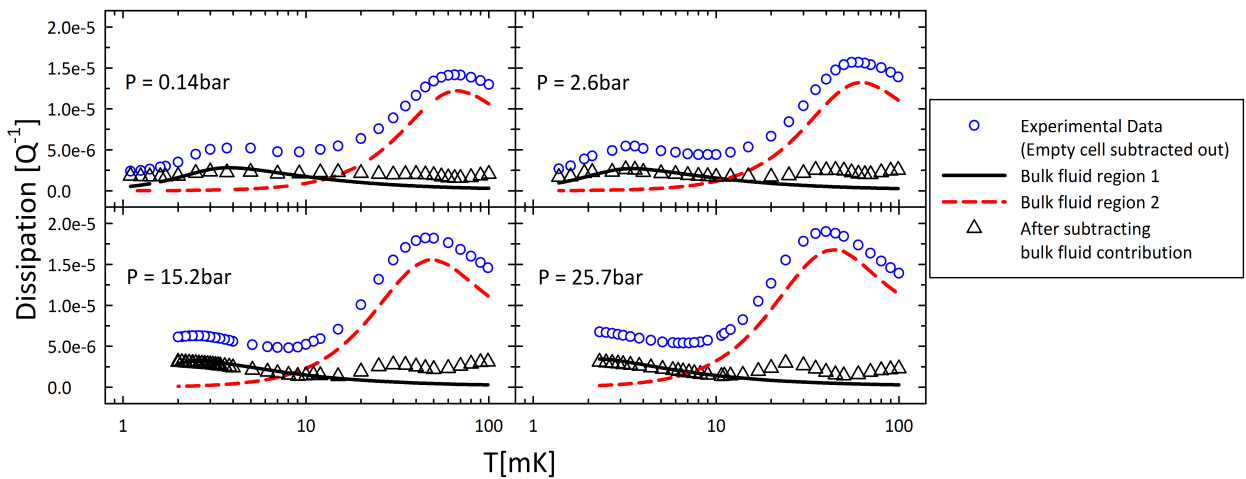}
\end{center}
\caption{(Color online) Experimental data for $Q^{-1}$ $vs$ temperature for four pressures after empty cell data is subtracted (open (blue) circles). Shown also are the fits for the bulk fluid contribution to the $Q^{-1}$ for two components - bulk fluid contained in the fill line (solid (black) line) and the bulk fluid possibly contained around the periphery of the cell, modeled as a channel of thickness 28 nm (dashed (red) line). After subtracting off the two bulk fluid contributions, the dissipation due to the $^3$He and aerogel combination is shown as the open (black) triangles. The dissipation of $\sim2.4\times10^{-6}$ is essentially temperature and pressure independent within the experimental error in the normal state data ($\sim \pm 0.6\times10^{-6}$). \label{NormDiss}}
\end{figure*}

\subsection{Normal State}

We obtained data for the resonant frequency, ($f(T)$), and dissipation factor $Q^{-1}(T)$, subtracting the values for the empty cell, at discrete temperatures between 100 mK and directly above the superfluid transition temperature with the cell being filled with liquid $^3$He at four different pressures: at the saturated vapor pressure (0.14 bar), 2.6, 15.7 and 25.7 bar. In order to minimize the thermal gradient between the experimental cell and the $^3$He melting curve thermometer (our primary thermometer), we waited for several hours after changing the temperature before recording each data point. We emphasize that the data for the empty cell has been subtracted from the data presented in this article.

\begin{figure*}
\begin{center}
\includegraphics[%
  width=6.8in,
  height = 4.7in]{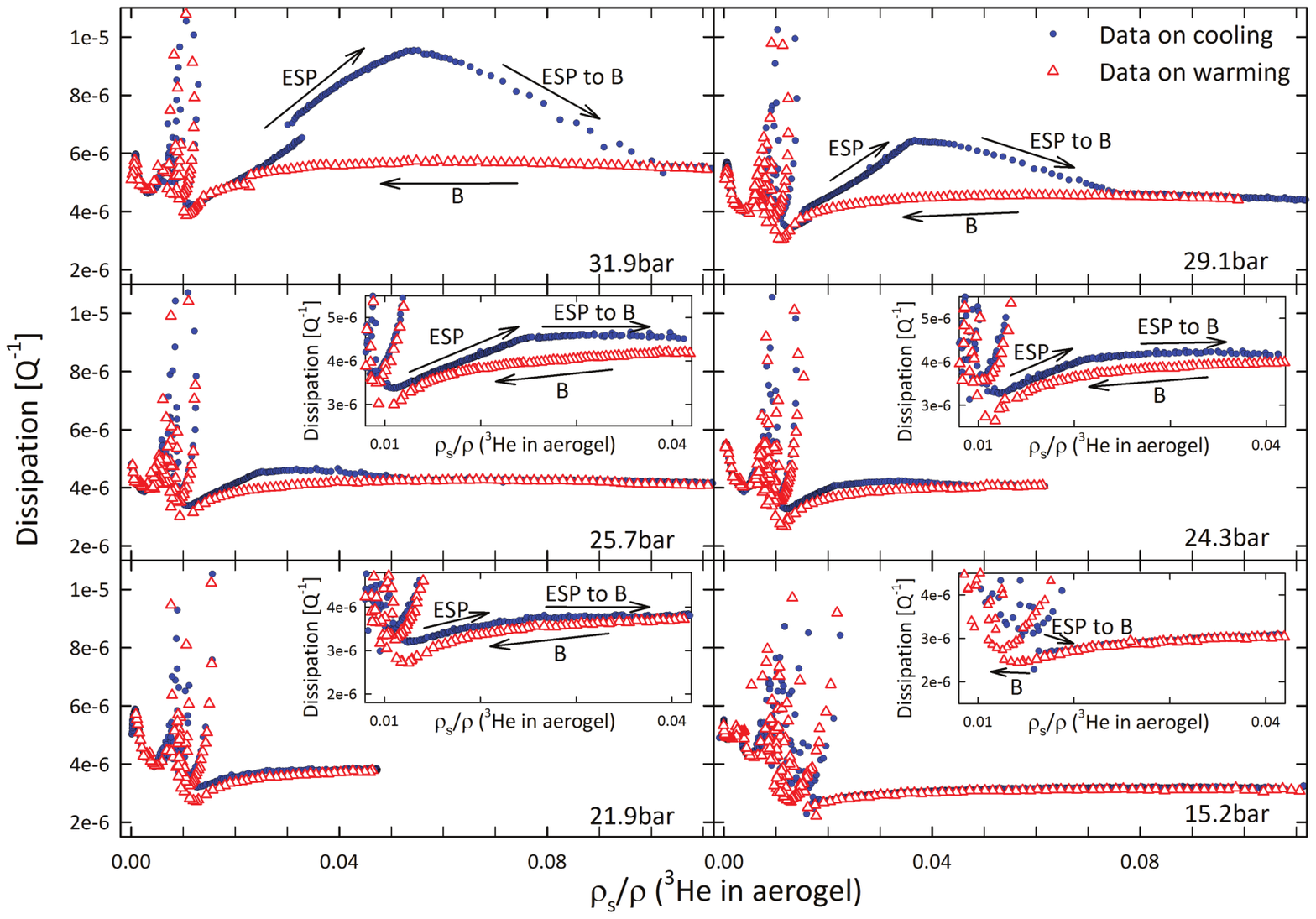}
\end{center}
\caption{(Color online) Data for $Q^{-1}$ vs $\rho_s/\rho$ at six different pressures is plotted. We note slow mode resonance crossings for $\rho_s/\rho < 0.015$. ESP-phase (cooling-blue solid circles) and B-phase (warming-red open triangles) coexistence regions are shown in the insets at lower pressures. We note the much larger dissipation in the superfluid ESP-phase than in the B-phase as well as the different functional dependence on the superfluid fraction for the two phases. Bulk fluid contributions have been subtracted, assuming bulk B phase. The discontinuity in the 31.9 bar data on cooling is due to the bulk A to B transition.\label{SuperfDissAll}}

\end{figure*}

\begin{figure*}
\begin{center}
\includegraphics[%
  width=7in,
  height = 2.9in]{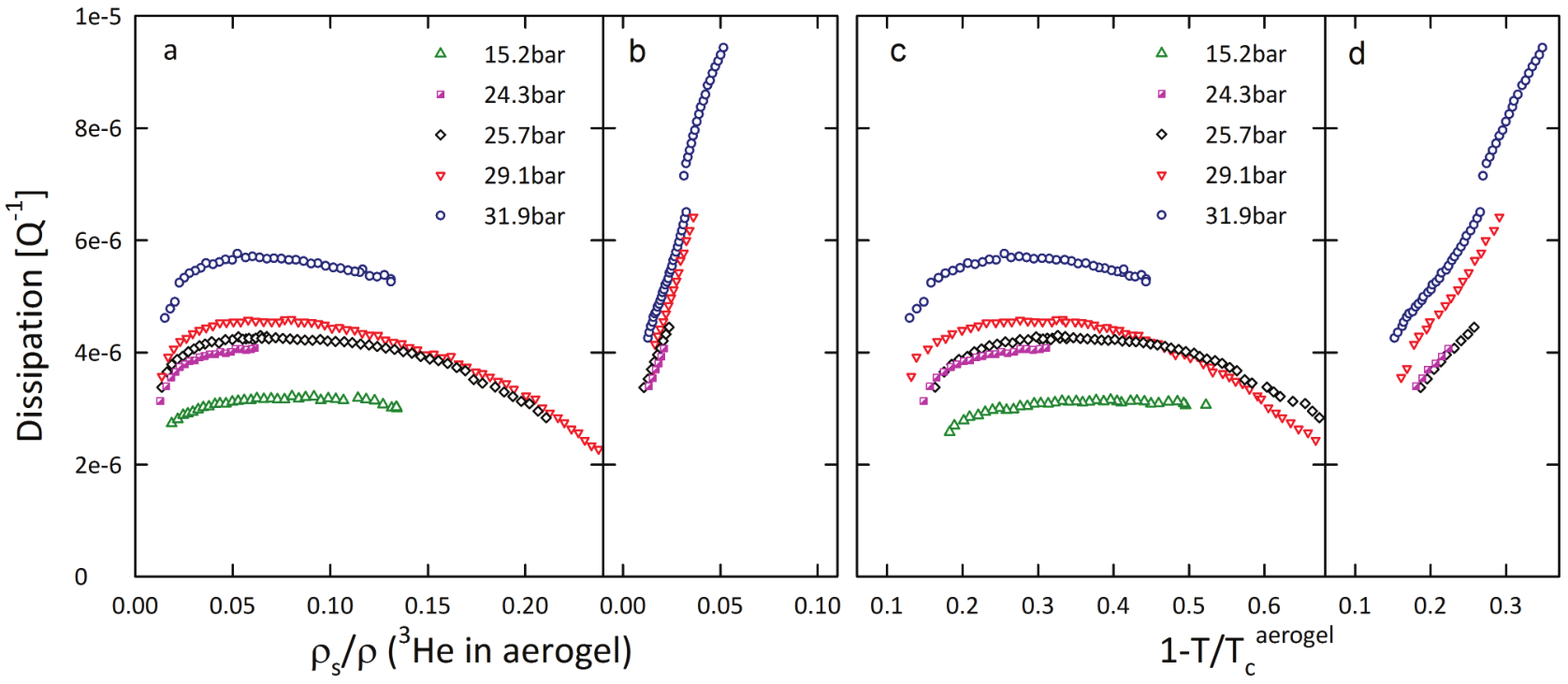}
\end{center}
\caption{(Color online) (a) A plot of $Q^{-1}$ in the B-phase vs $\rho_s/\rho$ for all the data sets in Fig. \ref{SuperfDissAll} combined in one plot. We note the significant pressure dependence of $Q^{-1}$ with $\rho_s/\rho$. Strong coupling effects seem to enhance the anomalous superfluid dissipation, which is in contrast to the normal state data, where we see no pressure dependence.\\
(b) A plot of $Q^{-1}$ in the ESP-phase vs $\rho_s/\rho$ for all the data. The pressure dependence of the dissipation is seen to arise mainly due to the larger extent in temperature of the ESP phase at high pressure;  the $Q^{-1}$ appears to scale as $\rho_s/\rho$. Discontinuities in the data are due to the bulk A $\rightarrow$ B transition on cooling.\\
(c) and (d)  Plots of $Q^{-1}$ vs $1-T/T_c^{aerogel}$ in the B-phase and ESP-phase respectively. Much of the scaling behavior is lost in this view compared to that seen in (a) and (b). \label{SuperfDissSep}}

\end{figure*}

Figure \ref{NormDiss} summarizes the data for the energy dissipation factor versus temperature in the normal state at four widely spaced experimental pressures.
The contribution from bulk fluid regions in the cell is shown by the solid and broken lines.
Data for the resonant frequency of the pendulum in the normal state was used to determine the fraction of moment of inertia of the bulk fluid regions.
Subtracting this contribution from the measured inertia and dissipation we observe a residual dissipation of $\sim (2.4\pm0.6)\times10^{-6}$ that we attribute to the $^3$He liquid in the aerogel.
Uncertainty in the estimation of the bulk fluid contribution arises from limitations in our model, as well as uncertainty in the inputs to that model such as the fluid viscosity. In addition, any residual temperature gradients between the thermometer and the cell will produce significant scatter in our data.
It is important to note that the dissipation does not have an obvious temperature dependence and any pressure dependence cannot be discerned from the plot in Fig. \ref{NormDiss}.

\subsection{Superfluid State of $^3$He in Aerogel}

Data was taken on both cooling and warming in the superfluid state at a number of different pressures.
Thermal gradient error could be rendered negligible in these measurements by maintaining a constant cooling (warming) rate ($\sim 30$ $\mu$K/hr) and through the observation of the bulk $T_c$ value, which is precisely known.

The fraction of the superfluid in the aerogel, ($\rho_s$/$\rho)_{aerogel}$ can be determined through the period shift of the pendulum. We calculate what its period would be if all the fluid were fully locked ($p_0$). With knowledge of the empty cell period, $p_{empty}$, we can define the decoupled fluid fraction as $[p_0-p(T)]/[p_0-p_{empty}]$, where $p(T)$ is the inverse of the measured resonance frequency.
Finally, we calculate and subtract the contribution to the decoupled fluid fraction of the bulk fluid regions in the cell. The remaining fraction of uncoupled moment of inertia in the cell is directly proportional to ($\rho_s$/$\rho)_{aerogel}$.

The calculated fractions due to the dissipation factor from the bulk fluid are also subtracted from the dissipation factor data we acquired below the superfluid transition. For $Q^{-1}$, the bulk fluid contribution is at most $4\times 10^{-6}$.
A summary of the data for  $Q^{-1}$ vs ($\rho_s$/$\rho)_{aerogel}$ at six different pressures is shown in Fig. \ref{SuperfDissAll}.

When ($\rho_s$/$\rho)_{aerogel}$ $\approx$ 0.015, the evolving velocity of the slow mode\cite{GolovSound} (a fourth sound like mode in which the superfluid oscillates with respect to the normal fluid which is clamped to the non-rigid aerogel) generates standing waves in the cavity resulting in a series of resonant modes whose frequencies cross the torsion oscillator frequency. In Fig. \ref{SuperfDissAll} we can identify these resonance peaks as a number of closely spaced ``loops."
The loop trajectory in $\rho_s/\rho$ vs $Q^{-1}$ is due to the extra dissipation and resonant frequency ``pulling" effect associated with resonance mode crossing as we drive the system with a phase locked loop.
These resonance effects\cite{GolovSound,nazaretski,McKenna,Goldner,Hook} in this region will be ignored in our subsequent discussions.

The superfluid transition and phase diagram for this sample were identified in our previous publication.\cite{BennettZhelev} Below the superfluid transition, we enter the superfluid ESP-phase on cooling.
At lower temperatures we observe a continuous phase transition between the ESP and the B phase (extended over a temperature interval of $\sim70 \mu$K).
It is thought that this width is due to the strong pinning of the interface by the aerogel.
On warming we stay in the B-phase until just below the critical temperature.
The reappearance of the ESP phase is very pressure dependent.
This strongly hysteretic behavior allows us to probe ESP and B phase properties over an extended temperature window, especially at elevated pressures.

We can observe the pressure dependence of the torsion pendulum dissipation in Fig. \ref{SuperfDissSep} where the dissipation obtained at different pressures are plotted together, with the horizontal axes being $1-T/T_c$ and $(\rho_s/\rho)_{aerogel}$.

We observe a broad peak in the dissipation in the B phase (Fig. \ref{SuperfDissSep}(a)). Below $T_c$, the dissipation rises, even though the impurity limited viscosity should be essentially constant. By contrast, the dissipation in the ESP phase rises even faster than in the B phase and appears to be proportional to $\rho_s/\rho$ as the temperature is lowered (Fig. \ref{SuperfDissSep}(b)).
This is in sharp contrast with experiments in the bulk, where the expectation is for the viscosity to drop sharply below $T_c$ and scale as $e^{-\Delta/k_BT}$ in the finite size regime.\cite{einzelparpia86}
$Q^{-1}(T)$ scales well with $\rho_s/\rho$ and not $(1-T/T_c)$, as shown in Fig. \ref{SuperfDissSep}.
Since $\rho_s/\rho$ $\propto$ $\Delta^2$, this implies that $Q^{-1}$ and the energy gap $\Delta$ are related.
In particular, we see that the anomalous dissipation of the ESP phase scales almost linearly with $\rho_s/\rho$, and exceeds the corresponding value of $Q^{-1}$ at the same $\rho_s/\rho$ (and $T/T_c$) in the B phase.
As the pressure is increased, the ESP phase $Q^{-1}$ rises considerably above that of the B phase and is emphasized due to the increased width of the ESP phase with pressure.

\section{Collision Drag Model in a Torsion Pendulum Geometry}

A starting point in the model for the dynamics of the helium-aerogel system is to map out the velocity profiles of the fluid and the aerogel across the flow channel.
We expect the fluid to be in a Drude flow regime,\cite{einzel, Takeuchi} where velocity of the fluid with respect to the aerogel is constant across the channel, with the exception of a small region of size $\delta_d=\sqrt{(2\eta/\rho\tau_f)}$ away from the edges.\cite{Takeuchi, Obara}
The frictional relaxation time $\tau_f$ is related to the friction force coupling the helium with the aerogel matrix:\cite{higashitani, higashitaniprb, higashitanijltp}
\begin{equation}
\mathbf{F}(\mathbf{v_l},\mathbf{v_a})=\frac {\rho}{\tau_{ f}} (\mathbf{v_l}-\mathbf{v_a})\\
\end{equation}
where $\mathbf{v_l}$ and $\mathbf{v_a}$ are respectively the velocities of the helium liquid and the aerogel.
The frictional force can be related to the average change of momentum a quasiparticle experiences upon scattering from an aerogel impurity. It can be shown that:\cite{higashitaniprb}
\begin{equation}
\tau_f=\frac{\tilde{\tau}}{(1-\frac{\rho_s^0}{\rho})(1+\frac{F^1_s}{3}(1-\frac{\rho_s^0}{\rho}))}
\label{TauTemp}
\end{equation}
where $F^1_s$ is a Fermi-liquid Landau parameter and $\rho_s^0$ is the bare superfluid fraction, stripped of Fermi-liquid effects. The bare superfluid fraction is related to the measured superfluid fraction $\rho_s$ through:
\begin{equation}
1-\frac{\rho_s}{\rho}=\frac{m^{*}}{m}\frac{1-\frac{\rho_s^0}{\rho}}{1+F^1_s(1-\frac{\rho_s^0}{\rho})}
\end{equation}

In the normal state, $\tilde{\tau}$ is just the transport relaxation time equal to the quasiparticle (qp) mean free path divided by the Fermi velocity. The quasi-particle mean free path can be estimated from the suppression of the superfluid transition as was discussed in Ref. \onlinecite{BennettZhelev} using a model first proposed by Abrikosov and Gorkov in Ref.\onlinecite{AG} and refined into the Isotropic Inhomogeneous Scattering Model (IISM) described in Ref. \onlinecite{thuneberg1} and Ref. \onlinecite{impurity,sands1,sands2}. The value of $\tau_f$ inferred from the 155 nm mean free path used to fit the $T_c$ suppression is $\approx 5 \times 10^{-9} s$, assuming a Fermi velocity of 30 m/s.

The sound velocity in the aerogel sample is expected to be in the range of $c\sim 30 - 50$ m/s;\cite{GolovSound, aerogelSound} for a frequency of 2.1 kHz we expect a compressional sound mode wavelength of a few millimeters. This is an order of magnitude larger than the relevant dimensions of the cell, yet there is a small but finite displacement of the aerogel strands relative to the motion of the adjacent wall's surfaces. The normal helium is well locked to the aerogel; the aerogel and helium form a composite medium exhibiting a velocity profile largely determined by the viscoelasticity of the aerogel.  Through numerical calculations, we expect about 1\% difference in the velocity in the middle of the cell and the wall. This velocity profile gives rise to a significant contribution to the dissipation in the cell. In addition, there is a small velocity difference between the entrained fluid and the aerogel itself that arises due to the finite value of $\tau_f$.

To solve for the velocity profiles of the helium and the aerogel, we write the Navier-Stokes and wave equations, coupled by the collision drag force:
\begin{eqnarray}
\rho\dot{\Omega}_{l}=\eta\frac{\partial^{2}\Omega_{l}}{\partial z^{2}}-\frac{\rho}{\tau_{f}}(\Omega_{l}-\Omega_{a}) \label{PDE1}\\
\rho_{a}\dot{\Omega}_{a}=i\frac{\mu}{\omega}\frac{\partial^{2}\Omega_{a}}{\partial z^{2}}+\frac{\rho}{\tau_{f}}(\Omega_{l}-\Omega_{a}) \label{PDE2}
\end{eqnarray}
where $\Omega_l(z)$ and $\Omega_a(z)$ are the angular velocity profiles of the helium liquid and the aerogel across the channel.
The shear modulus of the aerogel is $\mu$, which we can deduce from the aerogel sound velocity.
The viscosity of the helium $\eta$ would be similar to that of the bulk at high temperatures, but reach an impurity limited value at about 10 mK.
As such, we expect $\eta \lesssim 0.01$ Poise.

Once we have solved for the velocity profiles, we find the induced torque on the walls of the cell.
We assume that the velocity of both the liquid and the aerogel is equal to the velocity of the cell wall, i.e. $\Omega_{l(a)}(\pm z/2)=\dot{\theta}$.
\begin{equation*}
N_{l}=-\pi R^4 \eta \left( \frac{\partial\Omega_{l}}{\partial z} \right) _{z=h/2} \hspace{0.06in}
N_{a}=-i \pi R^4 \frac{\mu}{\omega} \left( \frac{\partial\Omega_{a}}{\partial z} \right) _{z=h/2}
\end{equation*}

Empty cell dissipation measurements show a nonzero value, even when extrapolated to $T=0$.
Yet, a purely elastic aerogel should not be dissipative.
A previous iteration of this experiment used an aerogel sample (grown in a different process) with a height of $\approx$ 4 mm. The otherwise essentially identical torsion pendulum containing that cell had a $Q$ $\approx$ 100$\times$ lower than the one described here. We can expect a $h^2$ dependence of the dissipation, with $h$ being the height of the cell. Furthermore, there have been a number of experiments on silica aerogels (though on samples denser compared to ours and at room temperature \cite{Viscoelasticity, AerogelAttenuation1, AerogelAttenuation2}), that report a complex elastic modulus, which would lead to dissipation effects associated with the plastic deformations of the aerogel.
We write the shear modulus of the aerogel as $\mu = \mu_{re}-i\mu_{im}$.

Accounting for the complex shear modulus, we obtain:
\begin{gather}
Q^{-1}(T) = -\frac{\textrm{Re}(N_{a}+N_{l})}{I_{0}\omega\dot{\theta}} \nonumber \\
\approx \frac{I_{a}}{I_{0}}\left(1+\frac{\rho_n(T)}{\rho}\frac{\rho}{\rho_{a}}\right)^{2}\frac{\rho_a\omega^{3}h^{2}}{12\mu_{re}^2}\left(\eta(T)+\frac{\mu_{im}}{\omega}\right)\nonumber\\
 +\frac{\rho_n(T)}{\rho}\frac{I_{l}}{I_{0}}\omega\tau_{f}
\label{NormModel}
\end{gather}

More details about the exact solution to the equations of motion and how we derive the result for $Q^{-1}$ can be found in Appendix B.

\section{Discussion}

There are three terms in Eq. \ref{NormModel} that contribute to the normal state dissipation.
The first one is proportional to the normal fluid viscosity $\eta(T)$, and is due to the aerogel flexure modifying the velocity profile of the liquid and causing extra dissipation.
Using $\eta \sim 0.01$ Poise, this term accounts for a contribution to $Q^{-1}$ of the order of $\sim 10^{-8}$. In order to match the experimental value of $Q^{-1}=2.4 \times 10^{-6}$, we need $\eta$ to be two orders of magnitude larger, which we consider unphysical.

The third term in Eq. \ref{NormModel} contains contributions to $Q^{-1}$ arising from the frictional relaxation time $\tau_f$. For this term to have a large enough contribution to match the experimental data for $Q^{-1}$, we need $\tau_f \sim 10^{-7}$ s. However, the quasiparticle mean free path in a 98\% open aerogel has been shown to be $\lesssim 200$ nm.\cite{lancaster_gapless, BennettZhelev, Choi_heatcap, lancasternormal, grenoble, dlca} Assuming a Fermi velocity of 30 m/s and effective mass $m^{*}/m \sim 3-5$, we find that $\tau_f$ above $T_c$ can at most be a few nanoseconds.

We suggest that the large temperature independent normal state dissipation could be due to the intrinsic dissipative nature of the aerogel, characterized by the ratio $\mu_{im}/\mu_{re}^2$.
The reason we are sensitive to the aerogel intrinsic dissipation term is the low resonant frequency of the torsion pendulum.
Since this term depends on $\mu_{im}/\omega$, its contribution would be less significant at the higher frequencies employed in previous ultrasound attenuation experiments reported in Ref. \onlinecite{Takeuchi} and \onlinecite{ChoiHigaSound}.
To obtain $Q^{-1}$ of the order of $10^{-6}$, we need $\mu_{im}/\mu_{re}\sim0.1$. Such a large loss tangent could be due to the fractal nature of the aerogel or could be related to the expected presence of a few monolayers of solid $^3$He on the surface of the aerogel strands.

\begin{figure}[floatfix]
\begin{center}
\includegraphics[%
  width=3.4in,
  height = 2.1in]{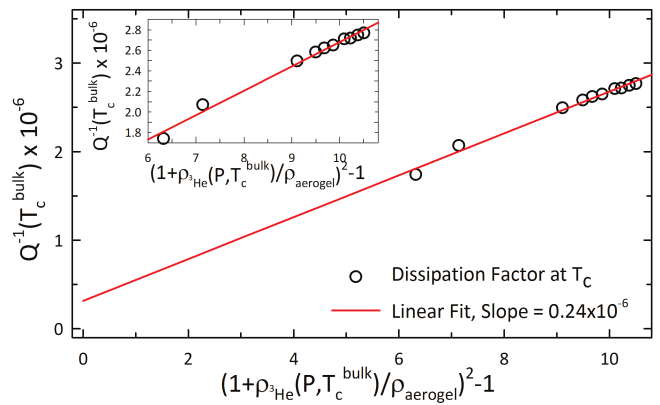}
\end{center}
\caption{(Color online) A plot for the dissipation measured at $T_c$ for 0.14, 3, 15.2, 18.5, 20.1, 21.9, 24.3, 25.7, 27.5 and 29.1 bars (experimental pressure increases as we go from left to right) with the bulk fluid and empty cell dissipation subtracted versus $(1+\rho_s/\rho_{aerogel})^2-1$. A linear regression line is shown, with a slope of $~2.4\times10^{-7}$ and y-intercept of $~3.1\times10^{-7}$. \label{TcDiss}}

\end{figure}

Figure \ref{TcDiss} shows the values for $Q^{-1}$ at $T_c^{bulk}$ with the bulk regions and empty cell contributions subtracted plotted versus $(1+\rho(P, T_{c}^{bulk})/\rho_{a})^2-1$ for a number of experimental pressures.
The transition temperature as a function of pressure is well known as are the density and the viscosity of the bulk fluid at $T_c$.
This allows us to accurately subtract the bulk fluid contributions and reveal the subtle pressure dependence of the data in the normal state.
Note that if our assumption that the main contribution to the dissipation in the normal state comes from the lossy aerogel, we would expect a linear relationship. A linear fit to the data is shown in Fig. \ref{TcDiss}, providing an evidence in support of this model. The y-intercept of $~3.1\times10^{-7}$ could be due to the uncertainty of the empty cell data (due to thermal lag between the empty cell and the Melting Curve Thermometer).

Assuming that energy dissipation of the torsion pendulum due to the interaction of the normal state excitations and the aerogel scales as $\left[1+\left(1-\frac{\rho_s}{\rho}\right)\frac{\rho}{\rho_a}\right]^2$, then such a contribution will decrease as the cell is cooled below T$_c$ and deeper in the superfluid state.
This cannot explain the dissipation we see in both the ESP and B phase superfluid states.

\begin{figure}[t]
\begin{center}
\includegraphics[%
  width=3.3in,
  height = 2.1in]{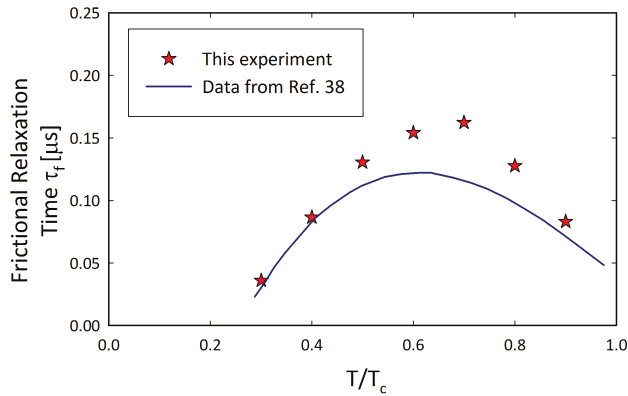}
\end{center}
\caption{(Color online) A plot for $\tau_f$ as a function of temperature assuming that the $\tau_f$ term in Eq. \ref{NormModel} is responsible for all of the extra dissipation we observe in the B phase. The data plotted is for 29.1 bar. We also show the data from Ref. \onlinecite{Obara}, which is deduced in a similar way.}
\label{SuperfTauf}
\end{figure}

We compare the broad peak in dissipation we observe in the B phase and find it to be similar to the sound attenuation data below $T_c$ observed in Ref. \onlinecite{ChoiHigaSound} and in Ref. \onlinecite {Obara} (Fig. \ref{SuperfTauf}).
We subtract the normal fluid contribution (using parameters from the linear fit in Fig. \ref{TcDiss}) and consider the residual dissipation.
If we allow its origin to be due to the $\frac{\rho_n(T)}{\rho}\frac{I_{l}}{I_{0}}\omega\tau_{f}(T)$ term, we can plot the so-inferred $\tau_f(T)$ as a function of the temperature.
Fig. \ref{SuperfTauf} shows this for the 29.1 bar data along with the data from Ref. \onlinecite{Obara}.
We find a good agreement between the two experiments, implying that the observed dissipation in the 50 kHz sound attenuation experiment and the torsion pendulum $Q^{-1}$ in the superfluid B phase probably have a similar origin. The relaxation time $\tilde{\tau}$ in Eq. \ref{TauTemp} can be shown to increase as we enter the superfluid state due to the rapid opening of the superfluid gap, before $\tilde{\tau}$ eventually diminishes to near zero at extremely low temperatures.\cite{higashitaniprb}
In addition, the denominator in Eq. \ref{TauTemp} should also decrease as $\rho_s/\rho$ grows.
These two effects combined produce a temperature dependence of $\tau_f$ with a similar shape to what we observe in Fig. \ref{SuperfTauf}.
We could expect an enhancement of $\tau_f$ in the superfluid state up to a factor of ten compared to its value at $T_c$. However, in order to produce a peak $\tau_f$ of order 0.15 $\mu$s, we need $\tau_f(T_c) \gtrsim 10$ ns, a value which is higher than the few nanoseconds that would be consistent with the $T_c$ suppression measurements.
Thus temperature variation of the frictional relaxation time cannot solely produce the observed data. Therefore, we conclude that there is an additional mechanism to dissipate energy not captured in the collision-drag model presented in Section IV and related to the emergence of the superfluid order.

One established way superfluid currents can dissipate energy is through interactions with bound states pinned to the boundary with the normal fluid at the vortex cores.\cite{Kopnin} This leads to a mutual friction term which can be shown to be proportional to\cite{Hall} $$\frac{\rho_s}{\rho}\frac{\rho_n}{\rho}(\bf{v}_s-\bf{v}_n)$$ Such a term would produce a peak in the dissipation similar to what we observe in our data for the B superfluid phase. No evidence for vortex states has been found in our experiment. The velocity amplitude for the superfluid current is small, much smaller than the velocities the fluid is driven at in typical experiments observing vorticity.\cite{Hall, Reppy} We also do not detect a noticeable increase in $Q^{-1}$ as we drive the pendulum harder. While the vortex dynamics model may not be applicable to our experiment, one can imagine that regions of normal fluid with the size of a typical vortex core (coherence length) exist, bound to denser regions of the aerogel. Such bound states will allow for lower energy excitations to interact with the superfluid flow and provide a mechanism for energy dissipation.

An object (in this case an aerogel strand) moving through bulk superfluid with velocity $v$ should feel a force that scales as $e^{-\Delta/k_bT}v$, as shown in Ref. \onlinecite{Fisher_Diss}. Assuming that the nodes of the ESP state order parameter tend to orient in the plane of the flow, then we would expect that the ESP state should be associated with higher dissipation than the B phase. However, this argument doesn't explain the different functional dependence of the dissipation in the ESP phase in terms of $\rho_s/\rho$ compared to that of the B phase.

Finally, we note that the pressure dependence of the observed dissipation in terms of pressure could be related to the degree of gap suppression in both ESP and B superfluid phases. Dissipation is higher at high pressures, where the gap suppression is less severe, and lower at lower pressures where the superfluid gap tends to be less pronounced and the density of states at lower energies increases.

\section{Conclusion}
We presented torsion pendulum $Q^{-1}$ data for a compressed aerogel sample filled with $^3$He in both normal and superfluid states. We developed a model for the normal fluid dynamics as embedded in the viscoelastic aerogel. We assert that frictional relaxation time is not large enough to account for either normal or superfluid $Q^{-1}$ data. Instead, we propose that dissipation features of the data below the superfluid transition originate from the superfluid state.

\begin{acknowledgments}
We thank S. Higashitani and J. Sauls for fruitful discussions. We acknowledge support from the NSF under DMR-1202991 at Cornell and DMR-1103625 at Northwestern.
\end{acknowledgments}

\appendix

\section{Bulk fluid contribution}

\begin{figure}[t]
\begin{center}
\includegraphics[
  width=3.2in,
  height = 4.85in]{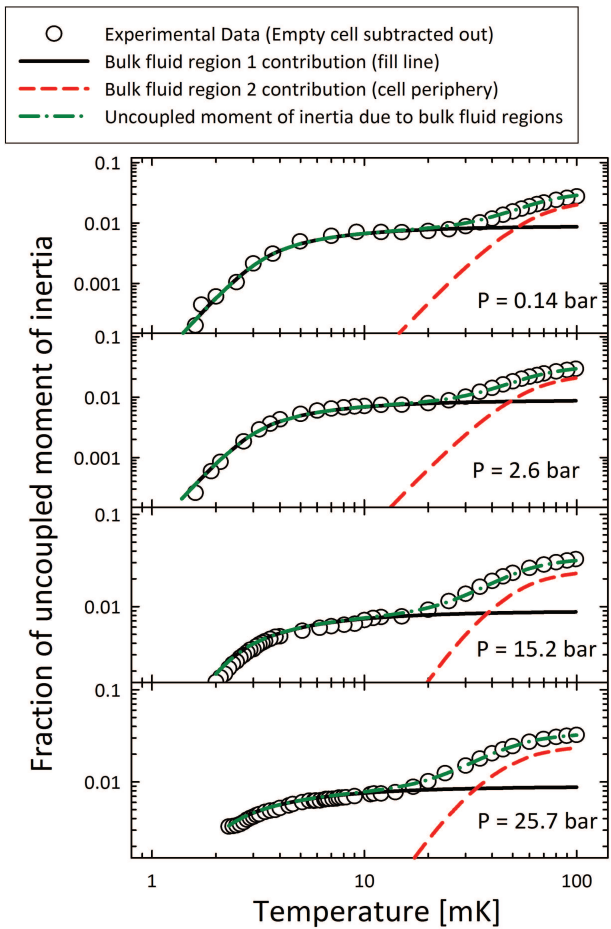}
\end{center}
\caption{(Color online) The fraction of fluid decoupled from the pendulum $vs$ temperature for four pressures after background subtraction (open circles). Also shown are the fits for the bulk fluid contribution for two components - Region 1, fluid in the fill line, a 1 mm diameter, 6 mm long cylinder comprising 0.8\% of the total fluid moment of inertia (solid (black) line), and Region 2, fluid at the periphery of the cell, modeled as a cavity of height 28 $\mu$m (dashed (red) line) comprising 3.2\% of the moment of inertia. \label{UncoupledFract}}
\end{figure}

We expect the normal state helium liquid to be well locked to the strands of the aerogel.
In the normal state any change in the resonant frequency compared to that of a cell with a fully locked fluid should originate from the bulk-like fluid regions of the cell. Figure \ref{UncoupledFract} shows data for the fraction of the moment of inertia not coupled to the walls of the cell at the four experimental pressures that were shown in Fig. \ref{NormDiss} (0.14, 2.6, 15.2 and 25.7 bar). The decoupled fluid fraction and dissipation show temperature dependent behavior characteristic of two distinct bulk fluid regions (two peaks in the normal state dissipation data, two ``shoulders" in the normal state decoupled fraction data).

The effective length of the fill line in the torsion rod and the cast epoxy cell is 6 mm long and 1 mm in diameter. The resulting bulk fluid column amounts to 0.8\% of the moment of inertia of the fluid in the cell. The fluid in the fill line is designated as bulk fluid Region 1.

In order to calculate the contribution to dissipation and period shift coming from the fluid in the fill line, we start by calculating the angular velocity profile $\Omega_\theta(r)$ by using the Navier-Stokes equation in a tall cylindrical geometry, which leads to
\begin{equation}
\frac{\partial^2 \Omega}{\partial r^2} + \frac{3}{r} \frac{\partial \Omega}{\partial r} + \frac{i \omega \rho}{\eta} \Omega = 0
\label{DiffEq}
\end{equation}
with $\Omega(\text{radius of the cylinder}) = \Omega_{cell}$.

Solving for $\Omega$ we can find the torque exerted by the fluid:
\begin{equation}
N = 2 \pi R^3 h \eta \left( \frac{\partial \Omega}{\partial r}\right)_{r=R}= \beta_1+i \omega \beta_2
\label{Torque}
\end{equation}
where $\beta_1$ contributes to the damping of the pendulum and $\beta_2$ to the moment of inertia. Temperature dependence of these values is determined by the temperature dependence of the viscosity of the fluid, $\eta(T)$.

\begin{figure}
\begin{center}
\includegraphics[
  width=3.3in,
  height = 3.2in]{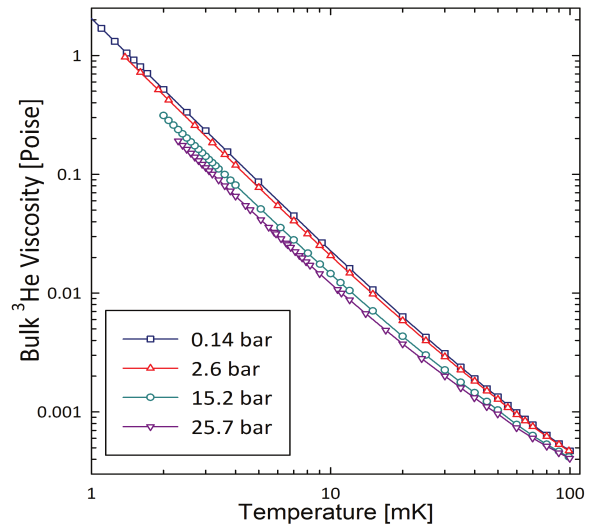}
\end{center}
\caption{(Color online) Values of viscosity in the normal state at the four experimental pressures.\label{Visc}}
\end{figure}

Near $T_c$ we expect normal state bulk viscosity to scale as $T^{-2}$.
At $T>10$mK the viscosity starts deviating from the Fermi liquid theory $T^{-2}$ behavior.
In order to calculate the higher temperature values, we use the following relations between the thermal conductivity ($\kappa$), heat capacity ($C_V$) and the viscosity:
\begin{align}
\kappa &=  \frac{1}{3}C_V v_F^2\tau_\kappa \\
\eta &= \frac{1}{5}\frac{m^*}
{m}\rho v_F^2\tau_\eta \\
C_V &= m^* \frac{\pi^2 k_B}{\hbar^2}(\frac{V}{3\pi^2 N})^{2/3} RT
\end{align}
Assuming that density and molar volume do not change in the temperature range 1-100 mK, and assuming $\tau_\eta \propto \tau_\kappa$,  we can infer that $\eta \propto \kappa$.
To find the exact values for the viscosity in the normal state, we use the values for $\eta(T_c)$ given in Ref. \onlinecite{archieJLTP} and Ref. \onlinecite{parpiaPRL}, and $\kappa(T_c)$ in Ref. \onlinecite{greywallkappa} and divide the two values to find the proportionality factor. We then multiply $\kappa(T)$  from  Ref. \onlinecite{greywallkappa} by this factor for each of the pressures we are interested and we find $\eta(T)$ up to 100 mK. The values for the viscosity for the four experimental pressures we used to calculate bulk fluid contribution in the normal state are shown in Fig. \ref{Visc}
In the superfluid state, experimental values for the superfluid fraction are taken from Ref. \onlinecite{parpiajltp} and for the viscosity from Ref. \onlinecite{einzelparpia86}

Numerically solving Eq. \ref{DiffEq}, we can calculate the contribution from the bulk fluid in the fill line. This contribution is shown with a solid (black) line in Fig. \ref{NormDiss} and \ref{UncoupledFract}. It is evident in Fig. \ref{UncoupledFract} that there is bulk fluid within the cell we have not yet accounted for.

The steel cavity containing the aerogel was dry fitted in the epoxy cast to prevent epoxy running in. We believe this resulted in small pockets of bulk fluid existing around the periphery of the cell. While we cannot do an exact calculation for the effects of these regions the same way as we did for the fluid in the torsion rod, we can still use the uncoupled moment of inertia data (Fig. \ref{UncoupledFract}) to estimate the contribution to the pendulum's dissipation. We assume that the relationship between the real and the imaginary part of the torque arising from the cell periphery bulk fluid is the same as that of a uniform thickness film encompassing all of the cell. For a thin film of fluid with a thickness $h$ and inertial contribution $I_{per}$, the torque exerted is $N=\beta_1+i \omega \beta_2$, with:
\begin{eqnarray}
\beta_1=\omega I_{per} \frac{\delta}{h} \frac{\sin(h/\delta)-\sinh(h/\delta)}{\cos(h/\delta)+\cosh(h/\delta)} \\
\beta_2=I_{per} \frac{\delta}{h} \frac{\sin(h/\delta)+\sinh(h/\delta)}{\cos(h/\delta)+\cosh(h/\delta)} \vspace{10pt}
\end{eqnarray}
where $\delta = \sqrt{2\eta/\rho\omega}$ is the viscous penetration depth of the fluid. Fitting to the dissipation data in Fig. \ref{NormDiss}, we find $h=28$ $\mu$m and $I_{per} = 0.032 I_f$, where $I_f$ is the moment of inertia of all the helium in the torsion pendulum head. These values are consistent with our expectations. The accuracy to which the epoxy cast and stainless steel cell are machined is within one-thousand of an inch, i.e. $25$ $\mu$m, and a film of that thickness around all of the cell surface amounts to $~0.05 I_f$. Since the bulk fluid is more likely coming from a few separate regions around the periphery, rather than from a continuous film, we would expect that $I_{per}\lesssim 0.05 I_f$. We also use these values and the viscosity of $^3$He to obtain the fraction of decoupled fluid from the periphery (Region 2) which we plot as the dashed (red) line in Fig. 6.

At the lowest experimental pressures (0.14, 2.6 and 4 bar), the liquid in the aerogel does not transition to a superfluid state.
At these pressures, the resonance period shift below $T_c$ originates from the bulk fluid regions.
In addition to the bulk fluid decoupling, we observe fourth sound resonance crossings effects, which occur at specific values of the sound velocity and therefore $\rho_s/\rho^{bulk}$. We obtain a good fit to these data using the model described in this appendix, which gives an independent confirmation that bulk fluid effects are fully accounted for.
More information about these effects can be found in the supplementary material of  Ref. \onlinecite{BennettZhelev}.

\section{Dynamics of normal $^3$He in aerogel}

We start by rewriting equations \ref{PDE1} and \ref{PDE2} as:
\begin{align}
\frac{\partial^2 \Omega_a}{\partial z^2} + a_a \Omega_a - b_a \Omega_l &= 0 \\
\frac{\partial^2 \Omega_l}{\partial z^2} + a_l \Omega_l - b_l \Omega_a &= 0
\end{align}
where we have defined the coefficients $a$ and $b$ as:
\begin{eqnarray}
a_a = i\frac{\rho \omega}{\mu} \hspace{0.17in} b_a = a_a\left(1-i \omega \tau_F \frac{\rho_a}{\rho}\right)\\
a_l = -\frac{\rho}{\eta \tau_F} \hspace{0.4in} b_l = a_l\left(1-i \omega \tau_F \right)
\end{eqnarray}

Solving the coupled differential equations, we arrive at:
\begin{widetext}
\begin{eqnarray}
\Omega_a \left(z\right) = \left[ \frac{D-\frac{a_l-a_a}{2}-b_a}{2D} \frac{\cos(k_1 z)}{\cos(k_1 h/2)}+\frac{D+\frac{a_l-a_a}{2}+b_a}{2D} \frac{\cos(k_2 z)}{\cos(k_2 h/2)}\right]\dot{\theta}     \label{vel_a}\\
\Omega_l \left(z\right) = \left[ \frac{D-\frac{a_a-a_l}{2}-b_l}{2D} \frac{\cos(k_1 z)}{\cos(k_1 h/2)}+\frac{D+\frac{a_a-a_l}{2}+b_l}{2D} \frac{\cos(k_2 z)}{\cos(k_2 h/2)}\right]\dot{\theta}     \label{vel_l}
\end{eqnarray}
\end{widetext}
where $D = \sqrt{\left(\frac{a_l-a_a}{2}\right)^2+b_l b_a}$ and $k_{1,2}=\sqrt{\frac{a_l+a_a}{2} \pm D}$.

To obtain a qualitative picture of the velocity profiles, we can explore the fact that $\omega \tau_f \ll 1$ and $\eta\omega/\mu \ll 1$. The coefficients $\left(D \pm \frac{a_l-a_a}{2} \pm b_{a,l}\right)/2D$ and the values of $k_{1,2}$ in Eq. \ref{vel_a}, \ref{vel_l} are approximated to the lowest order. This approximation gives us the $z$ dependence of the angular velocity of the aerogel, $\Omega_a(z)$, and that of the fluid $\Omega_l(z)$:
\begin{align}
\Omega_a(z) &\approx \dot{\theta} \frac{\cos{\left( \sqrt{\frac{\left(\rho+\rho_a \right)\omega^2}{\mu}}z \right)}}{\cos{\left(  \sqrt{\frac{\left(\rho+\rho_a \right)\omega^2}{\mu}}\frac{h}{2} \right)}} \label{vel_a_approx}\\
\Omega_l(z) &\approx \dot{\theta}\left[\frac{\cos{\left( \sqrt{\frac{\left(\rho+\rho_a \right)\omega^2}{\mu}}z \right)}}{\cos{\left(  \sqrt{\frac{\left(\rho+\rho_a \right)\omega^2}{\mu}}\frac{h}{2}\right)}}-i\omega\tau_f\frac{\cosh{\left( \frac{z}{\delta_d} \right)}}{\cosh{\left(  \frac{h}{2\delta_d}\right)}}\right] \label{vel_l_approx}
\end{align}
where $\delta_d = \sqrt{\eta\tau_f/\rho} \ll h$ is the ``dirty" fluid penetration depth, i.e. the length scale over which the velocity of the helium fluid deviates from the Drude flow regime with respect to the aerogel velocity. We observe that the shape of both aerogel and fluid velocity profiles is largely set by the elastic modulus of the aerogel, $\mu$. The relative velocity difference between the aerogel and the helium fluid is of the order of $\omega \tau_f \ll 1$ of the total velocity.


Eq. \ref{vel_a_approx} and \ref{vel_l_approx} present a qualitative picture for the differences in the velocities of the flow and the aerogel, but we need to include higher order terms in the expressions above to estimate the dissipation factors associated with the aerogel and the fluid in the cell. Importantly, we also allow the possibility of the elastic modulus of the aerogel to be a complex number, $\mu = \mu_{re}-i\mu_{im}$, with $\mu_{im}/\mu_{re}\ll1$. Then for $k_{1,2}$ we have:
\begin{align}
k_1 &\approx \sqrt{\frac{\left(\rho+\rho_a \right)\omega^2}{\mu_{re}}}\left[1+i\left(\frac{\omega\eta}{2\mu_{re}}+\frac{\mu_{im}}{\mu_{re}}+\frac{\omega\tau_f}{2} \frac{\rho}{(\rho+\rho_a)}\right)\right]\\
k_2 &\approx \frac{i}{\delta_d}\left[1-i\left(\frac{\omega\eta}{2\mu_{re}}+\frac{\omega\tau_f}{2}\right)\right]
\end{align}
As for the coefficients in \ref{vel_a}, \ref{vel_l}:
\begin{align}
C_1&=\frac{D-\frac{a_l-a_a}{2}-b_a}{2D}\approx 1\\
C_2&=\frac{D+\frac{a_l-a_a}{2}+b_a}{2D}\approx \left(\omega\tau_f\right)\left(\frac{\eta\omega}{\mu_{re}}\right)\\
C_3&=\frac{D-\frac{a_a-a_l}{2}-b_l}{2D}\approx 1+i\omega\tau_f\\
C_4&=\frac{D+\frac{a_a-a_l}{2}+b_l}{2D}\approx -i\omega\tau_f\left(1+i\frac{\eta\omega}{\mu_{re}}\frac{\rho+\rho_a}{\rho}\right)
\end{align}
The expressions for the induced torque by the aerogel ($N_a$) and the helium liquid ($N_l$) can be written as:
\begin{align}
N_a&=i\omega I_a \frac{2\mu}{\rho_a\omega^2 h}\left[C_1 k_1 \tan\left(k_1 \frac{h}{2}\right) + C_2 k_2 \tan\left(k_2 \frac{h}{2}\right)\right]\dot{\theta}\\
N_l&=I_l \frac{2\eta}{\rho h}\left[C_3 k_1 \tan\left(k_1 \frac{h}{2}\right) + C_4 k_2 \tan\left(k_2 \frac{h}{2}\right)\right]\dot{\theta}
\end{align}
Further, the expressions for the tangents can be approximated as:
\begin{align}
\tan&\left(k_1 \frac{h}{2}\right)\approx k_1 \frac{h}{2} + \frac{\left(\frac{\left(\rho+\rho_a \right)\omega^2 h}{4\mu_{re}}\right)^{3/2}}{3-\left(\frac{\left(\rho+\rho_a \right)\omega^2 h}{4\mu_{re}}\right)} +\\
&+i\frac{\left(\frac{\left(\rho+\rho_a \right)\omega^2 h}{4\mu_{re}}\right)^{3/2}}{1-\left(\frac{\left(\rho+\rho_a \right)\omega^2 h}{4\mu_{re}}\right)}\left( \frac{\eta\omega}{\mu_{re}}+\frac{\mu_{im}}{\mu_{re}}\frac{\omega\tau_f}{2} \frac{\rho}{\rho+\rho_a} \right) \nonumber\\
\tan&\left(k_2 \frac{h}{2}\right)\approx i
\end{align}
where we used the following relation:
\begin{equation}
\tan(\alpha+i\beta)\approx \alpha\left(1+\frac{\alpha^2}{3-3\alpha^2}\right)+i\beta\left(\frac{\alpha^2}{1-\alpha^2}\right)
\end{equation}
which is true in the case of $\beta \ll \alpha$ and $\alpha \lesssim 0.1$.
For the expression for $\tan(k_2 h/2)$, we use the fact that $\lvert k_2 h/2 \rvert \sim h/\delta_d \gg 1$ and that $\text{Im}(k_2 h/2)\gg \text{Re}(k_2 h/2)$.

Putting all of these expressions together, we arrive at:
\begin{align}
&N_{ind} = N_a+N_l\approx \\
&-\left[(I_a+I_l)\omega\frac{\xi}{3-\xi}\left(\frac{\eta\omega}{\mu_{re}}+\frac{\mu_{im}}{\mu_{re}}\right)+I_l\omega^2\tau_f\frac{3-\xi}{3-3\xi}\right]\dot{\theta} \nonumber\\
&+i\omega\frac{3-2\xi}{3-3\xi}(I_a+I_l)\dot{\theta} \nonumber
\end{align}
where $\xi = \frac{\left(\rho+\rho_a \right)\omega^2 h^2}{4\mu_{re}}$ and we have ignored terms containing $\delta_d/h\ll1$. We can simplify further, since $\xi\sim 10^{-2}$:
\begin{align}
N_{ind}&\approx -\left[(I_a+I_l)\frac{\left(\rho+\rho_a \right)\omega^4 h^2}{12\mu_{re}^2}\left(\eta+\frac{\mu_{im}}{\omega}\right)+I_l\omega^2\tau_f\right]\dot{\theta} \nonumber \\
&+i\omega(I_a+I_l)\dot{\theta}
\end{align}
Using this expression for the induced torque, we arrive at the expression for $Q^{-1}$ in Eq. \ref{NormModel}

\end{document}